\documentstyle[12pt,worldsci,epsfig]{article}
\def\photonatomright{\begin{picture}(3,1.5)(0,0)
                                \put(0,-0.75){\tencircw \symbol{2}}
                                \put(1.5,-0.75){\tencircw \symbol{1}}
                                \put(1.5,0.75){\tencircw \symbol{3}}
                                \put(3,0.75){\tencircw \symbol{0}}
                      \end{picture}
                     }

\def\photonatomup{\begin{picture}(1.5,3)(0,0)
                             \put(-0.75,3){\tencircw \symbol{3}}
                             \put(-0.75,1.5){\tencircw \symbol{2}}
                             \put(0.75,1.5){\tencircw \symbol{0}}
                             \put(0.75,0){\tencircw \symbol{1}}
                   \end{picture}
                  }

\def\photonrighthalf{\begin{picture}(30,1.5)(0,0)
                     \multiput(0,0)(3,0){5}{\photonatomright}
                  \end{picture}
                 }

\def\photonup{\begin{picture}(1.5,30)(0,0)
                  \multiput(0,0)(0,3){10}{\photonatomup}
               \end{picture}
              }

\def\fermionurr{\begin{picture}(30,15)(0,0)
                        \put(-30,-15){\vector(2,1){15}}
                        \put(-15,-7.5){\line(2,1){15}}
                  \end{picture}
                 }

\def\fermiondrr{\begin{picture}(30,15)(0,0)
                        \put(0,0){\vector(2,-1){15}}
                        \put(15,-7.5){\line(2,-1){15}}
                  \end{picture}
                 }

\newenvironment{Feynman}[3]{\begin{center}
                            \setlength{\unitlength}{#3 mm}
                            \begin{picture}(#1)(#2)
                            \thicklines
                           }{\end{picture} \end{center}}

\def\MW{$\mathrm{M}_{\mathrm{W}}$}

\newcommand{\mr}{\mathrm}

\newcommand {\oa} {\mbox{${\cal O}(\alpha)$}}

\newcommand{\bq}{\begin{equation}}
\newcommand{\eq}{\end{equation}}
\newcommand{\ba}{\begin{eqnarray}}
\newcommand{\ea}{\end{eqnarray}}

\newcommand{\nc}{\newcommand}
 \nc{\mb}[1]{\makebox[#1]{}}
 \nc{\CC}{{\scriptscriptstyle CC}}
 \nc{\NC}{{\scriptscriptstyle NC}}
 \nc{\V}{{\rm v}}
 \nc{\W}{{\scriptscriptstyle W}}
 \nc{\X}{{\scriptscriptstyle X}}
 \nc{\Z}{{\scriptscriptstyle Z}}
 
 \nc{\IE}{{\it i.e.,\ }}
 \nc{\EG}{{\it e.g.,\ }}
 \nc{\EA}{{\it et al.\ }}
 \nc{\AH}{{\it ad hoc\ }}
 \nc{\CHPT}{{$\chi_{\PT}$\ }}
\nc{\st}{\scriptstyle}
\nc{\sst}{\scriptscriptstyle}
\nc{\mco}{\multicolumn}
\nc{\vep}{\varepsilon}
\nc{\ra}{\rightarrow}
\nc{\nuN}{{\nu N_0}}
\nc{\nubN}{{\overline{\nu} N_0}}
\nc{\snuNC}{{\langle \sigma^{\nuN}_{\NC}\rangle }}
\nc{\snubNC}{{\langle \sigma^{\nubN}_{\NC}\rangle }}
\nc{\snuCC}{{\langle \sigma^{\nuN}_{\CC}\rangle }}
\nc{\snubCC}{{\langle \sigma^{\nubN}_{\CC}\rangle }}
\nc{\Rnu}{{R^{\nu}}}
\nc{\Rnub}{{R^{\overline{\nu}}}}
\nc{\sintW}{{\sin^2 \theta_{\W} }}
\nc{\vp}{{\bf p}}
\nc{\rz}{{\rho_0^2}}
\nc{\ko}{K^0}
\nc{\kb}{\bar{K^0}}
\nc{\al}{\alpha}
\nc{\ab}{\bar{\alpha}}

\begin{document}                                                                

\author{\ Arif~Akhundov \\
\it Institute of Physics, Azerbaijan Academy of Sciences, \\
\it H. Cavid ave. 33, Baku 370143, Azerbaijan}

\begin{center}
{\it Contribution to the NuFact08, 10th International Workshop on Neutrino Factories, Superbeams and Betabeams, Valencia, Spain, 30 June - 5 July 2008.}
\end{center}
                                                           
\title{\bf Applicability of the formulae of Bardin and Dokuchaeva for the  
       radiative corrections analysis in the NuTeV experiment}


\maketitle                                                                      
\setlength{\baselineskip}{2.6ex}

 \begin{abstract} 
We point out one of the possible sources of the "NuTeV anomaly": 
the effect of the non-adequate application of the one-loop electroweak radiative corrections including QED hard photon emission derived by Bardin and Dokuchaeva (1986) in the NuTeV radiative corrections data analysis of deep inelastic neutrino and anti-neutrino deep inelastic scattering. 
\end{abstract}

The NuTeV collaboration~\cite{NuTeV} has made a precise determination of the weak mixing angle by measuring charged and neutral current cross sections from neutrino and anti-neutrino deep inelastic scattering (DIS) on iron. Their value differs by 3 standard deviations from that obtained from measurements at the Z pole.

A precise determination of the on-shell weak mixing angle has been performed by the NuTeV collaboration for the first time through the measurements of the Pashos-Wolfenstein ratio~\cite{PW:73}:
\noindent 
\bq
 \mathrm R^{-} = \frac{\sigma(\nu_{\mu}N\rightarrow\nu_{\mu}X)-
                   \sigma(\bar\nu_{\mu}N\rightarrow\bar\nu_{\mu}X)}
                  {\sigma(\nu_{\mu}N\rightarrow\mu^-X)-  
                   \sigma(\bar\nu_{\mu}N\rightarrow\mu^+X)} .  
\eq
\noindent 

The NuTeV collaboration finds $\sintW=0.2277 \pm 0.0016$ which is 
3.0 $\sigma$ higher than that obtained from the Standard Model predictions. 

From this experimental value one obtains the mass of \MW~boson
~\cite{NuTeV}
\noindent 
\bq
\mathrm{M_W} = 80.14 \pm 0.08 \; \mathrm{GeV}\,
\eq
\noindent 
which is smaller than other measurements of~\MW~at LEP/SLD and the
Tevatron (Fig. 1).
\begin{figure}
\begin{center}
  \includegraphics[height=.3\textheight]{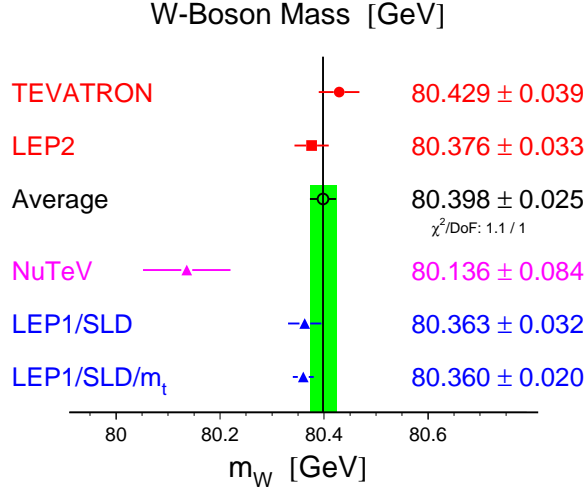}
\end{center}
 \caption{The results of the direct measurements of~\MW~at 
 LEP2/Tevatron are compared with the indirect determinations at LEP1/SLD 
 and in the NuTeV experiment~\cite{LEPEWWG,LEPEWWG_SLD:06,NuTeV}.}
\end{figure}

The radiative corrections (RC) are important for the higher statistics
experiments and dependent of the methods used to extract the wanted
cross section from the data~\cite{Madrid,HERMES}.
 
Here we point out one of the possible sources of the "NuTeV anomaly":
the effect of the non-adequate application of the Fortran program NUDIS~\cite{Bardin:86} 
for the calculations of the electroweak RC to the inclusive cross section of deep inelastic ${\nu_\mu}(\bar{\nu_{\mu}}) N$-scattering in the data analysis of neutrino and anti-neutrino DIS in the NuTeV experiment~\cite{NuTeV}.
This effect we consider as the most promising effect~\cite{Akhundov:07} that might reconcile the NuTeV measurement with the precise measurements near the $Z$ pole.

The RC produce a shift of the extracted on-shell weak mixing angle $\sin^2\theta_W$~\cite{Marciano:80,Bardin:86,Hollik:04}
\noindent 
\bq
\Delta\sin^2\theta_W =
\frac{\frac{1}{2}-\sin^2\theta_W+\frac{20}{27}\sin^4\theta_W}
{1-\frac{40}{27}\sin^2\theta_W}
\left( \delta R^\nu_\NC + \delta R^\nu_\CC \right), 
\eq
\noindent 
where $\delta R^\nu_\NC + \delta R^\nu_\CC $ is the total electroweak RC
~\cite{Bardin:86} to $R^\nu$~\cite{LlewellynSmith}  
\noindent 
\bq
R^\nu = \frac{\sigma_\NC^\nu(\nu_\mu N\to\nu_\mu X)}
{\sigma_\CC^\nu(\nu_\mu N\to\mu^- X)},
\eq
\noindent
and 
\bq
\delta R^\nu_{NC} = \frac{\sigma^{\mathrm{Corr.}}_{\nu NC}
-\sigma^{\mathrm{Born}}_{\nu NC}}{\sigma^{\mathrm{Born}}_{\nu NC}}\, ,
\qquad 
\delta R^\nu_{CC} = - \frac{\sigma^{\mathrm{Corr.}}_{\nu CC}
-\sigma^{\mathrm{Born}}_{\nu CC}}{\sigma^{\mathrm{Born}}_{\nu CC}}\, 
\eq
are the corrections to the NC and CC cross sections.

The main contribution to the total RC arises from $\delta R^\nu_\CC$
~\cite{Bardin:86,Hollik:04,Bardin:05,Hollik:05}, i.e. from the charged current events in the neutrino DIS: 
\noindent 
\bq
\nu_\mu(k_1) + N(p_1) \to \mu^-(k_2) + X(p_2). 
\eq
\noindent 

The electroweak RC to DIS have two different parts - weak RC and QED RC.  The contribution of the weak RC does not depend from the event selection in the experiment, but the contribution of the QED RC depends significantly on the measured kinematical quantities. The contribution of the QED RC is a complicated function of kinematical variables used in the cross section 
measurement: the radiative corrections calculated in a different set of variables can have a completely different value and behavior~\cite{Akhundov:96,Akhundov:04} because of the different bremsstrahlung contribution to (6) from the process:
\noindent 
\bq
\nu_\mu(k_1) + N(p_1) \rightarrow  \mu(k_2)+ X(p_2) + \gamma(k).
\label{brems}
\eq
\noindent 
with non-observed photon(s).

If the initial energy of neutrino $E_{\nu}$ in the lab. frame is known the fixed target experiments on the neutrino DIS could use only two additional experimentally measured quantities to determine the kinematics of the events.  

By measuring the energy $E_{\mu}$ and the angle $\theta_{\mu}$ of the scattered charged lepton the analysis of deep inelastic events will then be based on the evaluation of, the so-called invariant leptonic variables:
\noindent 
\bq
Q_l^2 = -(k_1 - k_2)^2, \hspace{1.cm} y_l = \frac{p_1 (k_1 - k_2)}{p_1 k_1},
\hspace{1.cm} x_l =  \frac{Q_l^2}{S y_l}
\label{qxyl}
\eq
\noindent 
with
\noindent 
\bq
     S = (k_1 + p_1)^2 ~\simeq~2M E_\nu ,
\label{s}
\eq
\noindent 
where $M$ is the mass of nucleon.

In the same manner, by measuring the energy $E_{h}$ and the angle $\theta_{h}$ of the hadron jet the analysis of deep inelastic events could be based on the evaluation of, the so-called invariant hadronic variables~\cite{Akhundov:96}:
\noindent 
\bq
 Q_h^2 = -(p_2 - p_1)^2,\hspace{1.cm} y_h = \frac{p_1 (p_2 - p_1)}{p_1 k_1},
\hspace{1.cm} x_h = \frac{Q_h^2}{S y_h}.
\label{qxyh}
\eq
\noindent 

In the composition of both measurements there are many possible sets of variables. One set is, the so-called invariant mixed variables:
\noindent 
\bq
 Q_m^2 = Q_l^2, \hspace{1.cm} y_h = \frac{p_1 Q_h }{p_1 k_1},
\hspace{1.cm} x_m = \frac{Q_l^2}{S y_h}.
\label{qxym}
\eq
\noindent   

The general formula~\cite{Akhundov:96} for the radiatively corrected neutrino DIS cross section in terms of leptonic variables can be represented as the sum of the Born distribution with the contributions due to virtual loop diagrams and real hard photon emission:
\noindent 
\bq
  \frac{d^2\sigma^{RC}}{dxdQ^2}  =
  \frac{d^2\sigma^{Born}}{dxdQ^2}~(1~+~\delta^{V}(x,Q^2))+
 \int\int{dx_hdQ_h^2}~H(x,Q^2,x_h,Q_h^2)~\frac{d^2\sigma^{Born}}{dx_hdQ_h^2}
\label{RC}
\eq
\noindent 

 The part of (\ref{RC}), proportional to $\delta^{V}(x,Q^2)$, contains
 the contributions from the EW and QED loop corrections and
 from the soft part of the real photon radiation. 
 The second part accounts for the bremsstrahlung contribution (7) where the function $H(x,Q^2,x_h,Q_h^2)$ is the hard photon radiator. 

The explicit formulae for $\delta^{V}(x,Q^2)$ and $H(x,Q^2,x_h,Q_h^2)$ are  
derived in the unpublished communication~\cite{Bardin:86} in the framework of the quark-parton model and in the approximation of the four-momentum contact interaction neglecting the terms of the order 
$\alpha Q^2/M_{W}^2$. Moreover, in~\cite{Bardin:86} for  
the density function of the initial quark in the nucleon 
$f_i(x,Q)$ the scaling approximation is used which simplifies the calculation of the twofold integral in (\ref{RC}).

 It is worth to note, that the twofold integral in (\ref{RC})
 depends on the structure functions of the nucleon, not only at a given 
 $ F(x,Q^2) $ point,
 but in the physical region of $(x_h,Q_h^2)$ defined by the kinematics
 of the process (\ref{brems}) in leptonic variables~\cite{Akhundov:96}\\ 
 (Fig. 2.)\footnote{The kinematical boundaries 
 ${\mathrm Q^2}_{h \mathrm I,II}$ are defined  
 by formula (B.17) of ~\cite{Akhundov:96}.}.

\begin{figure}[bhtp]
\begin{center}
\mbox{
\epsfysize=9.cm
\epsffile[0 0 530 530]{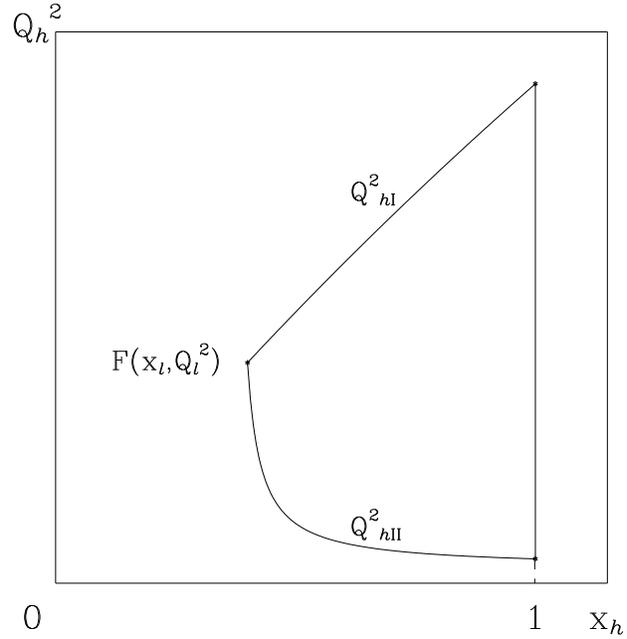}
}
\end{center}
\caption{
Integration region of $(x_h,Q_h^2)$ for the DIS cross section in leptonic
variables.
\label{intxq}
}
\end{figure}
 

The integration region of $(x_h,Q_h^2)$ accounts for the contribution of the radiative process (see Fig. 3) to the inclusive cross section of the CC and NC DIS and is described by the Bjorken variables:
\noindent 
\bq
      x~\leq~x_h~\leq~1,\hspace{1.cm} 0~\leq~y_h~\leq~1,\hspace{1.cm}
      M^2~\leq~M_h^2~\leq~W^2 ,
\label{xyM}
\eq
\noindent 
where the invariant masses $M_{h}^{2}$ and $W^2$ of the hadronic final 
state are defined as
\noindent 
\bq
   M_{h}^{2} = M^2 + Q_h^2  \left( \frac{1-x_{h}}{x_{h}} \right),
   \hspace{1.cm} W^2 = M^2 + Q^2_l \left( \frac{1-x_l}{x_l} \right). 
\eq
\noindent 
\setlength{\textwidth}{168mm}
\setlength{\textheight}{240mm}
\setlength{\oddsidemargin}{-0.3cm}

\vspace*{1.0cm}
\begin{minipage}[t]{7.8cm}{
\begin{center}
\begin{Feynman}{60,60}{0,0}{0.8}
%
\put(-11,62){$\nu_{\mu} (\vec k_1)$}  
\put( 50,62){$\mu^-(\vec k_2,m_{\mu})$}
\put(-09,-05){$N \, (\vec p_1,M)$}
\put( 50,-06){$X(\vec p_2,W)$}
\put(34,30){$Q_l^2$}
\put(00,60){\fermiondrr}
\put(60,60){\fermionurr}
\put(30.5,15){\photonup}
\put(30,15){\circle*{5}}
\put(30,17){\line(2,-1){30}}
\put(30,13){\line(2,-1){28}}
\put(30,45){\circle*{20.0}}
\put(30,15){\fermionurr}
\put(30,15){\fermiondrr}
\end{Feynman}
\vspace*{1.cm}
    Virtual loop diagrams
\end{center}
}\end{minipage}
\begin{minipage}[t]{7.8cm} {
\begin{center}
\begin{Feynman}{60,60}{0,0}{0.8}
%
\put(-11,62){$\nu_{\mu} (\vec k_1)$}  
\put( 50,62){$\mu^-(\vec k_2,m_{\mu})$}
\put(-09,-05){$N \, (\vec p_1,M)$}
\put( 50,-06){$X(\vec p_2,M_h)$}
\put(34,30){$Q_l^2$}
\put(24.5,3.75){$\gamma \, (\vec k)$}
\put(00,60){\vector(2,-1){13.5}}
\put(11,54.5){\line(2,-1){19.}}
\put(7.0,3.75){\circle*{1.5}}
\put(7.0,3.75){\photonrighthalf}
\put(45,52.5){\line(2,1){15}}
\put(30,45){\vector(2,1){19.}}
%
\put(30,45){\circle*{1.5}}
\put(30.5,15){\photonup}
\put(30,15){\circle*{5}}
\put(30,17){\line(2,-1){30}}
\put(30,13){\line(2,-1){28}}
\put(30,15){\fermionurr}
\put(30,15){\fermiondrr}
\end{Feynman}
\vspace*{1.0cm}
 Initial quark radiation
\end{center}
}\end{minipage}

\vspace*{1.0cm}

\begin{minipage}[tbh]{7.8cm}{
\begin{center}
\vspace*{0.5cm}
\begin{Feynman}{60,60}{0,0}{0.8}
%
\put(-11,62){$\nu_{\mu} (\vec k_1)$}  
\put( 50,62){$\mu^-(\vec k_2,m_{\mu})$}
\put(-09,-05){$N \, (\vec p_1,M)$}
\put( 50,-06){$X(\vec p_2,M_h)$}
\put(34,30){$Q_l^2$}
\put(71,3.75){$\gamma \, (\vec k)$}
\put(00,60){\fermiondrr}
\put(60,60){\fermionurr}
\put(30.5,15){\photonup}
\put(30,15){\circle*{5}}
\put(30,17){\line(2,-1){30}}
\put(30,13){\line(2,-1){28}}
\put(30,45){\circle*{1.5}}
\put(30,15){\fermionurr}
\put(30,15){\fermiondrr}
\put(53.0,3.75){\photonrighthalf}
\put(53.0,3.75){\circle*{1.5}}
\end{Feynman}
\vspace*{1.0cm}
  Final quark radiation
\end{center}
}\end{minipage}
\begin{minipage}[tbh]{7.8cm} {
\begin{center}
\vspace*{0.5cm}
\begin{Feynman}{60,60}{0,0}{0.8}
%
\put(-11,62){$\nu_{\mu} (\vec k_1)$}  
\put( 50,62){$\mu^-(\vec k_2,m_{\mu})$}
\put(-09,-05){$N \, (\vec p_1,M)$}
\put( 50,-06){$X(\vec p_2,M_h)$}
\put(34,30){$Q_h^2$}
\put(71,56){$\gamma \, (\vec k)$}
\put(00,60){\fermiondrr}
\put(60,60){\fermionurr}
\put(30.5,15){\photonup}
\put(30,15){\circle*{5}}
\put(30,17){\line(2,-1){30}}
\put(30,13){\line(2,-1){28}}
\put(30,45){\circle*{1.5}}
\put(30,15){\fermionurr}
\put(30,15){\fermiondrr}
\put(53.0,56.25){\photonrighthalf}
\put(53.0,56.25){\circle*{1.5}}
\end{Feynman}
\vspace*{1.0cm}
  Muon radiation
\end{center}
}\end{minipage}
\vspace*{0.5cm}

Fig. 3. Virtual loop and bremsstrahlung diagrams contributing to  
        deep inelastic ${\nu_{\mu}}N$-scattering .
\vspace*{0.5cm}

From the studies of the RC for DIS ~\cite{Akhundov:96}
it is known that the large radiative corrections are due to the emission of hard photons (the twofold integral in (\ref{RC})). 

For the hard bremsstrahlung (\ref{brems}) the minimum of $Q_h^2$ is
\bq
       ({Q_h^2})_{min}~\simeq~ x_l^2 M_h^2.
\label{Qmin}
\eq

This formula shows that the calculation of the contribution from hard photon emission demands the knowledge of the nucleon structure functions in the region $Q^2 \rightarrow 0 $. 

The NuTeV experiment uses~\cite{Zeller:PhD} the computer program
ZFITTER~\cite{ZFITTER} for the calculation of the electroweak corrections
and the formulae of Bardin and Dokuchaeva~\cite{Bardin:86} implemented in the Fortran program NUDIS~\cite{Bardin:86} which contains the virtual loop corrections and the bremsstrahlung contribution (Fig. 3) in leptonic variables without applying a cut on photon kinematics\footnote{With the  
exception of a cut on the energy of final hadrons $E_h > 10~GeV $.}.

The semi-analytical program NUDIS calculates the RC factor of the order {\oa} to the inclusive differential cross section 
$d^2 \sigma /dxdQ^2$ of neutrino and anti-neutrino CC and NC DIS at fixed energy of the neutrino beam.

In reality, the initial energy of neutrino $E_\nu$ is measured
for {\it each selected event}.   
In the NuTeV~\cite{NuTeV:05} the three experimentally measured quantities are: $E_\mu$ and $\theta_\mu$, the energy and the scattering angle of the outgoing muon, and $E_{HAD}$, the energy deposited in the target calorimeter
which includes the energy of the hadronic final state $E_h$ and the energy of the emitted photon $E_\gamma$:  
\noindent 
\bq
    E_{HAD} = E_h + E_{\gamma}, 
\eq
\noindent 
Then for this event the initial neutrino energy $E_\nu$ is calculated by:
\noindent 
\bq
    E_\nu = E_\mu + E_{HAD} 
\eq
\noindent
 
The measurement of $E_{HAD}$ for the event selection~\cite{NuTeV:05} means the detection of real hard photons with the energy $E_\gamma > \bar {E_\gamma}$, where $\bar {E_\gamma}$ is the photonic calorimeter threshold. 

Therefore, the contribution of such hard photons to the inclusive cross section of DIS should be subtracted from the bremsstrahlung integral 
in (\ref{RC}). This implies the integration in (\ref{RC}) over 
the physical region $(x_h,Q_h^2)$ restricted by the following condition:
\noindent 
\bq
 Q_h^2/x_h \geq Q_l^2/x_l - 2 M \bar {E_\gamma}
\label{boundary}
\eq
\noindent 

This is the main point of the non-adequate application of the formulae of 
Bardin and Dokuchaeva and the Fortran program NUDIS in the radiative corrections analysis of the NuTeV experiment. 

We anticipate substantial change of the value of the radiative correction factor $\delta(x,Q^2)$:  
\noindent 
\ba
\delta({x,Q^2})=
 \frac{d^2{\sigma}^{\mr{RC}}/{dxdQ^2}}
      {d^2{\sigma}^{\mr{Born} } /{dxdQ^2}}-1 
\label{delta}
\ea
\noindent 
for the inclusive cross section of the CC and NC DIS and for 
the values of $\delta R^\nu_\CC$ and $\delta R^\nu_\NC$ by adequate calculation of the contribution of hard photon emission.   

The recent re-calculation of the electroweak RC to neutrino DIS
~\cite{Bardin:05} including higher order contributions and different scheme of the subtraction of the mass singularities is performed for the hadronic variables and is used in the experiment NOMAD~\cite{Astier:2003rj}.
 The size of the QED RC to the CC scattering in hadronic variables are much smaller than the corrections in leptonic variables~\cite{Bardin:05}. 

\section*{Acknowledgements}
 I would like to thank the organizers of the NuFact08 Workshop 
 for kind hospitality. I thank K.~McFarland, Y.~Hayato and A.~Blondel for  
 the discussions. I am grateful to P.~Hernandez and N.~Rius for support  
 and to A.~Shiekh for remarks.


\end{document}